# Everett's Multiverse and the World as Wavefunction

Paul Tappenden
paulpagetappenden@gmail.com

> There may be a way to maintain a monistic wavefunction ontology, but it is certainly not *trivial* to see what that way is.
>
> [28, p. 129, original emphasis]

**Abstract:** Everett suggested that there's no such thing as wavefunction collapse. He hypothesized that for an idealized spin measurement the apparatus evolves into a superposition on the pointer basis of two apparatuses, each displaying one of the two outcomes which are standardly thought of as alternatives. And as a result the observer 'splits' into two observers, each perceiving a different outcome.

There have been problems. Why the pointer basis? Decoherence is generally accepted by Everettian theorists to be the key to the right answer there. And in what sense is probability involved, when all possible outcomes occur? Everett's response to that problem was inadequate. A first attempt to find a different route to probability was introduce by Neil Graham in 1973 and the path from there has led to two distinct models of branching. I describe how the ideas have evolved and their relation to the concepts of uncertainty and objective probability. Then I describe the further problem of wavefunction monism, emphasized by Maudlin, and make a suggestion as to how it might be resolved.

**Keywords:** Everett interpretation; many worlds interpretation; Everettian quantum mechanics; quantum probability; wavefunction realism

## 1. Two Concepts of Branching

There are two ways to model branching, both of which can be illustrated by the waters of a river being distributed into the streams of an estuary. If water is thought of as a continuous substance, as many physicists still believed not much more than a century ago, the river splits into lesser streams. But if water is composed of classical particles, each particle follows a linear trajectory taking it on a unique path to the sea; no particle splits. It was clearly the former analogy which Hugh Everett III had in mind when he proposed what he called the *relative state* formulation of quantum mechanics. He wrote of observers splitting in quantum measurement contexts [1].

An observer about to make an idealized z-spin measurement on an x-spin-up particle is to be thought of as splitting into an observer perceiving the outcome $U$ (z-spin-up) and an observer perceiving the outcome $D$ (z-spin-down). On a stochastic interpretation of quantum mechanics, where exclusively one or the other outcome is thought to occur, the pre-measurement observer assigns probabilities for the later occurrence of U and D according the Born rule, where the probability of U plus the probability of D is equal to 1 for the idealized case. But how can an observer who expects to split assign probabilities to future observations? Jeffrey Barrett argues that it was dissatisfaction with Everett's attempted response to this problem which led Neill Graham to propose an alternative structure to the branching process [2].

I shall use the analogy of a bifurcating road to make the change of perspective vivid. For the idealized measurement there are just two branches, one for the outcome U and the other for the outcome D. Each branch has a quantum amplitude whose absolute square is commonly called the branch weight (W) where $W_U + W_D = 1$. For the road analogy W can be imagined as corresponding to width. Everett thought in terms of a road of unit width bifurcating into two roads of lesser widths which sum to unity. The observer can be thought of as travelling along the unit road and splitting in an amoeba-like way at the junction into an observer on the U road and an observer on the D road. Everett himself used the analogy of an amoeba in an unpublished manuscript [4]. As Barrett explains, Everett thought of an observer conducting a series of such measurements as repeatedly splitting and attempted to analyze the statistical recordings for the resulting set of observers in terms of a notion of typicality which would do the work of a standard probabilistic assessment of statistical evidence.

Graham sought to bring a probability more directly into the picture by modifying the concept of branching. Instead of our road bifurcating into two roads it bifurcates into two *sets* of roads in such a way that the *measure* of the U set of roads is $W_U$ and the measure of D set is $W_D$. We can now imagine that on making the measurement the observer splits into some number $N(W_U)$ of observers observing U and $N(W_D)$ of observers observing D [3]. The idea is blatantly *ad hoc* but that's not the only problem.

As Barrett points out, there's an implicit assumption of a principle of indifference: the idea that the original observer is equally likely to find theirself on any one of the plethora of post-measurement branches [2]. It's also puzzling why *the* pre-measurement observer should somehow expect to find themself on any one of the post-measurement branches. In what sense can any one of the downstream observers be identified with the upstream observer? An attempt to meet the latter problem was made by David Deutsch when he wrote:

> I propose a slight change in the Everett interpretation:
> *Axiom 8*. The world consists of a continuously infinite-measured set of universes. By a 'measured set' I mean a set together with a measure on that set. [5]

He continued as follows (by substituting {2} for Deutsch's term I have adapted the citation to make 'model measurement' refer to our idealized spin measurement):

> Each of these subsets [of the original set of universes], which I shall call a *branch*, consists of a continuous infinity of identical universes. During the model measurement, the

> world has initially only one branch and is partitioned into {2} branches. The branches play the same role as individual universes do in Everett's original version, but the probabilistic interpretation is now truly 'built in'. [5]

To continue with the road analogy, we now have a continuous infinity of roads running parallel, on each of which are isomorphic observers about to make 'identical' measurements. Like the particles of water which never split, no road bifurcates but when the parallel measurements occur each road does a dogleg turn so that the original set of roads partitions into a subset which constitutes the $U$ branch, measure $W_U$, and a subset which constitutes the $D$ branch, measure $W_D$.

In each of Deutsch's identical universes the result of the target measurement is destined to be either $U$ or $D$ but an observer is in principle unable to determine in advance whether s/he inhabits a universe where the result will be $U$ or a universe where the result will be $D$. Although he doesn't explicitly say so, Deutsch thus appeals to a notion of self-location ignorance to introduce pre-measurement uncertainty. Prior to the measurement an observer doesn't know which sort of identical universe s/he inhabits, one in which $U$ will occur or one in which $D$ will occur. That's why Deutsch sees his proposal as making a probabilistic interpretation of the branching process 'built in'.[1]

Hilary Greaves interpreted Deutsch's proposal in this way, writing:

> David Deutsch suggests that, to solve the incoherence problem, the ontology of the many-worlds interpretation needs to be supplemented. In addition to the quantum state of the universe, we are to postulate a continuously infinite set of universes, together with a preferred measure on that set. The measure is such that, when a measurement occurs, the proportion of universes in the original branch that end up on a given branch is given by the mod-squared measure of that branch. Observers can then be *uncertain* about which outcome will occur *in the universe they are in*.
>
> [6, §1.2, original emphasis]

Strictly speaking the term 'proportion' doesn't apply to Deutsch's *infinite* set of identical universe but Greaves here recognizes that Deutsch has attempted to resolve what she called the *incoherence problem* for Everettian theory, though she didn't embrace the proposal. What Greaves refers to as the incoherence problem is, as she put it:

> How can it make sense to talk of probabilities (other than 0 and 1) at all, since all 'possible' outcomes actually occur? [6, §1.1]

---

[1] Deutsch has confirmed that he was indeed thinking in terms of self-location uncertainty at the time. Private communication.

Prior to measurement, Deutsch's observer is ignorant as to which sort of identical universe s/he inhabits but, on the basis of knowing the upcoming branch weights, assigns a subjective probability equal to $W_U$ to being in a universe where the measurement outcome will be *U* and a subjective probability equal to $W_D$ to being in a universe where the measurement outcome will be *D*. Axiom 8 can be seen as a modification of Graham's proposal in order to introduce pre-measurement uncertainty via self-location ignorance and assumes that measures on an infinite set can be taken as a guide to subjective probability assignments. As with Graham's proposal, choice of measure, which Greaves refers to above as 'a preferred measure' is *ad hoc,* simply chosen to make probability assignments conform to the Born rule and there is also an implicit appeal to a principle of indifference.

Responding to this problem about the choice of a preferred measure led Deutsch to be the first to introduce decision theory to deriving the Born rule for an interpretation of quantum mechanics which takes processes such as our idealized measurement as involving branching rather than stochasticity [7][2]. David Wallace has since done considerable further work on Deutsch's original idea so that the line of reasoning has come to be known as the Deutsch-Wallace argument [8, pp. 160-189]. Tim Maudlin forcefully rejects it and there have been several other challenges [9].

The Deutsch-Wallace argument claims to show that a well informed observer pre-measurement should assign subjective probabilities equal to what they take the post-measurement branch weights to be. And note that this effectively makes Deutsch's Axiom 8 redundant. If the argument is good it applies just as well to Everett's original conception of branching as like the bifurcation of a road. It might be objected that for that original conception, where the observer must expect to split, there is no available concept of pre-measurement uncertainty but it's controversial whether such uncertainty is required for the Deutsch-Wallace argument to go through. Greaves has argued against any need for uncertainty in Everettian theory [10].

**2 Vaidmanian uncertainty**

Even if there's some controversy as whether a notion of uncertainty needs to be associated with the concept of probability nobody would deny that intuitively things look more difficult for Everettian theory if any appeal to uncertainty is absent. And we have seen that a quest for a role for uncertainty seems to have motivated the first departures from Everett's idea that for our idealized measurement observers split in two. The departure from Everett's original amoebal concept of branching is nascent in Graham's proposal and blatant for Deutsch's Axiom 8.

Lev Vaidman was the first to notice that there's a way to associate uncertainty with Everett's original concept of splitting. He suggested considering the situation of an observer post-measurement, pre-observation [12]. Like Deutsch, Vaidman used the concept of self-location uncertainty. For our idealized spin measurement setup, the well-informed post-measurement, pre-observation observer is interpreted by Vaidman as being uncertain as whether s/he's on the *U* branch or the *D* branch. In such a situation Vaidman presumed that the observer should assign a subjective probability equal to $W_U$ to being on the *U* branch and $W_D$ to being on the *D* branch, an assumption which Paul Tappenden later called the Born-Vaidman rule [11, p. 104].

---

[2] Deutsch has said that it was Brice DeWitt who brought his attention to the necessity of tackling the measure problem. Private communication.

David Albert objected that Vaidmanian uncertainty 'comes too late, if it comes at all' [13]. Here, it's worth considering what role post-measurement, pre-observation uncertainty plays in *stochastic* theory. From that point of view, our idealized spin measurement would normally be performed by an observer who perceives the outcome as soon as it occurs. And since the notion of pre-measurement uncertainty seems unproblematic for stochastic theory there's no motivation to consider what the subjective probability assignments of a post-measurement observer *would* be *if* s/he were ignorant of the outcome. If the pre-measurement observer were offered a wager on outcomes it would seem unproblematic that s/he should make judgments as to how to bet on the basis of pre-measurement uncertainty.

But clearly in the stochastic case it's *possible* for an observer to be ignorant of the outcome post-measurement and, equally clearly, in such a situation the observer would make *exactly the same* judgments as to how to bet post-measurement as they would have done pre-measurement. So if the post-measurement, pre-observation observer were offered the same wager but told that stakes had to be laid *before* measurement they would clearly *regret* not having laid a stake beforehand if they had not done so. Knowing that in advance, the pre-measurement observer would have reason to lay a stake pre-measurement. The implication is that in the stochastic case minimizing possible future regret is just as good a reason to act as is maximizing possible future gain. The same argument applies in the Vaidmanian case, as was pointed out by Tappenden [11, p. 116]. Albert's objection is thus fully met. The mere *possibility* of post-measurement, pre-observation uncertainty is enough to give good reason to act pre-measurement. That hasn't been noticed for stochastic theory but is true there too.

Prior to Tappenden's argument, Vaidman suggested a different way to explain pre-measurement betting in splitting scenarios. He argued that the pre-measurement observer should 'care' about the weights of post-measurement branches in a way which would explain rational betting behavior [14]. Greaves later developed the idea at some length independently [10, §2.3]. Tappenden's argument effectively demonstrates that what the pre-measurement observer should care about is the gambling judgment of a possible post-measurement, pre-observation observer.

A further problem has remained for the traditional Everettian concept of branching as splitting. For our idealized spin measurement the Vaidmanian observer clearly *must* assign subjective probabilities to going on to observe *U* and *D* equal to $W_U$ and $W_D$ just as an observer applying stochastic theory must assign subjective probabilities equal to the objective probabilities for the outcomes *U* and *D*. Yet on the splitting model, unmodified either *à la* Graham or *à la* Deutsch, there are just two post-measurement, pre-observation observers according to Vaidman, one on the *U* branch and one on the *D* branch. In which case each would seem to have to think that they are either the observer on the *U* branch or the observer on the *D* branch and can't tell which so either is just as likely as the other. So each post-measurement, pre-observation observer should assign subjective probabilities of 0.5 to each possibility, *irrespective* of the values of $W_U$ and $W_D$. In other words, the probability judgments of the Vaidmanian observer should be ruled by a principle of indifference, which makes no sense at all for Everettian theory. There are currently two arguments that the principle of indifference should be overruled by the post-measurement, pre-observation observer which I shall not attempt to assess here [15] [16]. They are currently the last word for the Vaidmanian approach to Everettian splitting. Let's now go on to the last word for the alternative model of branching introduced by Deutsch.

**3 Parallel universes again**

Deutsch has not given up on his interpretation of branching as the partitioning of a set of identical universes but has recognized that if multiple universes prior to partitioning are thought of as being qualitatively identical the question naturally arises as to why partitioning should arise at all. What could bring about differences between qualitatively identical universes other than stochastic processes taking place within each of them? Invoking stochastic processes here would make the whole scheme pointless since that was exactly what Everett was trying to do without. Deutsch has attempted to address this problem by using the concept of fungibility. He writes:

> It *is* consistent for two identical entities to become different under deterministic and symmetrical laws. But, for that to happen, they must be more than just exact images of each other: they must be *fungible*
>
> [17, p. 265, original emphasis]

He continues:
> It is not that they [identical universes] coincide *in* anything, such as an external space: they are not in space. An instance of space is part of each of them. That they 'coincide' means only that they are not separate in any way.
> It is hard to imagine perfectly identical things coinciding. For instance, as soon as you imagine just one of them, your imagination has already violated their fungibility. But, although imagination may baulk, reason does not.
>
> [17, p. 269, original emphasis]

The suggestion is that thinking of identical universes as like proverbial peas in a pod is too simplistic and they should rather be thought of as like euros. There is no sense in asking the bank to return the very same euros which one deposited. The coins are *not* fungible but the euros are. However, this suggestion seems to create a problem because an observer can presumably *indexically refer* to the spin measurement device s/he is about to activate. The idea is supposed to be that the observer is ignorant as to *which one* of an infinite set of devices it is. It is either a device which is destined to show *U* or a device which is destined to show *D*. But it's impossible to indexically refer to a particular one of a set of objects which are fungible. I can point to the coin which I deposit at the bank but not to the euro which is thereby added to my account. So Deutsch's introduction of the concept of fungibility seems to undermine the idea that one can point to a device which has a determinate future which it is in principle impossible to know in advance. That suggests that if it's assumed that an observer *can* indexically refer to a spin measurement device, something generally taken for granted, then *either* the device operates stochastically *or* it splits in the way Everett suggested. If indexical reference isn't available then the device cannot have a predetermined future which the observer is *in principle* unable to determine in advance, given *total* knowledge of the universe which they inhabit. The problem of indexical reference has also arisen for an alternative view of branching which bears some similarity to Deutsch's idea of a partitioning set of parallel universes.

**4 Overlap**

Independently of Deutsch, Simon Saunders and David Wallace have also sought to introduce pre-measurement uncertainty to Everettian theory via a concept of self-location ignorance [18]. Their proposal is based on David Lewis's analysis of personal fission scenarios introduced to philosophical discussion of personal identity, independent of concerns about physics [19]. In describing Lewis's work Saunders and Wallace write:

> The trick was to suppose in the face of branching, say into two, that *there are two persons present all along* – persons who initially overlap or coincide. This is equivalent to the stipulation that by 'person', roughly speaking, we mean a unique cradle-to-grave continuant, specifically a unique spacetime worm. As for the meaning of 'overlapping', there are plenty of homely analogies: the Chester A. Arthur Parkway, [Lewis] observed, overlaps with Route 137 for a brief stretch, but still there are two roads.
> [18, pp. 294-5, original emphasis]

Two different roads can have a strip of tarmac in common. Likewise, two distinct persons can have 'temporal parts' of their bodies in common. Transposed to the context of the idealized spin measurement, *there are two observers present all along*, observer$_U$ who observes the outcome *U* and observer$_D$ who observes the outcome *D*. Prior to the measurement those two observers have temporal parts of their bodies in common, like the two roads having a stretch of tarmac in common. Furthermore, unlike Lewis, Saunders and Wallace suggest that, 'it is at least somewhat natural to attribute two sets of thoughts to those persons' [18, p. 303].

In other words, if well informed prior to measurement, both observer$_U$ and observer$_D$ can think: 'I'm one of two types of person. I'm either a person who's going to observe *U* or I'm a person who's going to observe *D*, but I'm ignorant as to which'. This is somewhat like Deutsch's proposal [5], a difference being that Saunders and Wallace generate the idea of there being multiple observers prior to the measurement via Lewis's concept of overlap rather that Deutsch's Axiom 8 which posits non-overlapping 'parallel' universes.

Saunders' and Wallace's overlap proposal was queried by Tappenden who argued that the putative pre-measurement observer$_U$ and observer$_D$ would not each be able to make indexical reference to their own bodies since before measurement their bodies have their temporal parts in common, like the overlapping roads [20]. Responding to that criticism, Saunders and Wallace emphasized that their analysis depends on an idea from linguistics know as the principle of interpretive charity, opening their reply to Tappenden with a quote from the philosopher of language Donald Davidson [21].

To see the linguistic principle of charity in action, imagine taking a bird's eye view of the idealized spin measurement given the overlap interpretation of the setup. What is seen prior to the measurement can seem to be a single observer making a single utterance of 'I'm one of two types of person. I'm either a person who's going to observe *U* or I'm a person who's going to observe *D*, but I'm ignorant as to which'. If what is heard is interpreted as the single utterance of one person then it's false, because there is one person who is going to split, not two persons one of whom is going to observe *U* and the other *D*.

But a more 'charitable' interpretation of what's heard is apparently possible; charitable because it takes what's heard to be *true*. What's heard can be taken to be two distinct utterances by two distinct observers even though only a single sound is heard. Rather as if the word 'here' written on a strip of tarmac common to those two overlapping roads were taken to refer to a position on the Chester A. Arthur Parkway and also, separately, to a position on Route 137. According to Saunders and Wallace, interpreting 'I'm one of two types of person. I'm either a person who's going to observe *U* or I'm a person who's going to observe *D*, but I'm ignorant as to which' as true is important and an apparently metaphysical concern about indexical reference can be set aside for the sake of preserving a concept of pre-measurement uncertainty in the face of branching. As Saunders and Wallace put it:

> Tappenden, by contrast, does seem to be looking for deep metaphysical truths: truths to which we have no access except via our pre-scientific intuitions, yet which we can know so surely that they bear on our choice of scientific theory.
>
> (*ibid.*: 317)

Saunders' and Wallace's 'choice of scientific theory' here refers to their choice of the overlap interpretation of branching, where there are two observers present prior to the idealized spin measurement, rather than the splitting interpretation where there is just one observer who splits. But in what sense are those different interpretations 'scientific' rather than metaphysical? It would seem that no possible experiment could decide between them, given Everett's hypothesis that processes which have been thought of as stochastic in fact involve branching.

**5 Divergence**

Following the overlap proposal, Saunders has made an alternative suggestion a bit closer to Deutsch's original idea. He does not suggest that quantum mechanics requires a supplementary axiom, as Deutsch did, but rather that the formalism allows branching to be interpreted as the partitioning of a set 'worlds' which are qualitatively identical prior to branching (like Deutsch's 'identical universes') but which do not overlap in the sense of having temporal parts in common [22, pp. 196-200]. He refers to this view of the partitioning of a set of worlds as 'divergence' and writes:

> once stated in this way, the suspicion is that whether worlds in EQM [Everettian Quantum Mechanics] diverge or overlap is *underdetermined* by the mathematics. One can use either picture; they are better or worse adapted to different purposes.
> [22, p. 200]

Earlier, he writes:

> The worry is not that overlapping worlds are unintelligible or inconsistent; it is that they make nonsense of ordinary beliefs […] Diverging worlds, composed of objects and events that do not overlap (that are qualitatively but not numerically identical) do not suffer from this problem.
> [22, p. 197]

That makes Saunders' preference clear for a divergence model of branching, rather than an overlap model. Alastair Wilson has articulated Saunders' concept of a divergence model of branching in a different way, writing:

> The diverging picture arises from a non-standard interpretation of the consistent histories formalism. The projection operators which feature in the consistent histories formalism are normally interpreted as representing token property-instantiations. This allows that objects or events in two different histories can be numerically identical, resulting in a metaphysic of overlapping Everett worlds. But if the projection operators are instead interpreted as representing types of property-instantiations, then it becomes possible for events in distinct histories to only ever be qualitatively identical to one another, generating a metaphysic of diverging Everett worlds.
> [23, pp. 714-715]

Wilson has developed Saunders' divergence concept of Everettian branching into a version of what Lewis has called 'modal realism' which takes talk of possible worlds to refer to concretely existing worlds. As Lewis puts it:

> There are so many other worlds, in fact, that absolutely *every* way that a world could possibly be is a way that some world *is*.
> [24, p. 2, original emphases]

Wilson restricts Lewis's notion of possibility to what is physically possible according to quantum mechanics. The divergence interpretation of branching then naturally translates into a modal realism where an observer inhabits an actual world which is just one of many, concretely existing, physically possible worlds. To see how that works, consider Saunders' divergence analysis of the idealized spin measurement.

Prior to the measurement there are many observers in worlds which do not overlap. Those worlds have had qualitatively identical histories up until the measurement event, which involves corresponding measurement devices in each world. Each device, it is claimed, does not operate stochastically, rather it is determined in advance whether it will record $U$ or $D$ but it is in principle impossible for the observer in each world to predict what the outcome in that world will be. Each observer pre-measurement is thus subject to self-location uncertainty. Each well informed observer knows that s/he inhabits a world where the result will be exclusively either $U$ or $D$ but does not know which is the case.

Wilson, very plausibly, translates the statement by a pre-measurement observer of: 'I'm in a world where the measurement outcome is destined to be $U$ or $D$ but I don't know which' as '*Possibly* I'm in a world where the outcome is destined to be $U$ and *possibly* I'm in a world where the outcome is destined to be $D$ but I don't know what is *actually* the case'. Thus an observer pre-measurement, according to Wilson, should think of themselves as actually inhabiting one of a set of extant, currently qualitatively identical, 'possible' worlds. Wilson's thoughts on this are fully developed in [25].

Recall that I mentioned that Deutsch has addressed the question as to why a set of qualitatively identical worlds should diverge. If those worlds do not contain stochastic processes what reason could there be for them to diverge? And I indicated possible reasons for concern about the ability of an observer to make indexical reference to objects in their environment, including their own body, both on Deutsch's conception of identical universes which are 'fungible' and in the case of overlapping worlds. There must at least be some suspicion that a similar thought applies to the Saunders-Wilson concept of divergence. I quoted Wilson above suggesting that the divergence interpretation of branching arises out of understanding quantum-mechanical representations as referring to types of property instantiations rather than token property instantiations. That would seem to make Saunders-Wilson parallel worlds fungible in the sense which Deutsch has attributed to his identical universes. Can an observer in one of a number of fungible worlds indexically refer to objects in their environment? It seems that that question needs to be addressed for any scheme which interprets branching as the partitioning of linear histories.

## 6 The world as wavefunction

There is as yet no consensus amongst philosophers of physics that Everettian probability problem has been resolved, far from it. But, as we've seen, Everettian theorists have mounted some ingenious challenges to the idea that the problem is intractable. At the same time, amongst Everettian theorists there's no consensus as to how uncertainty fits into the picture. Struggling with that problem has spawned distinct and incompatible models for the metaphysics of branching and discussion is ongoing.

I now want to take a look at another problem for Everettian theory which has been waiting in the wings. The question of wavefunction monism; the idea that all of material existence just is the universal quantum wavefunction. In recent years there has been much discussion about the existential status of the wavefunction. What I want to do here is propose a metaphysics which may be appropriate for wavefunction monism in an Everettian context. We have seen how fundamental the role of metaphysics has been in the development of ideas about branching and uncertainty. Perhaps metaphysics has a fundamental role to play as well in the debate about the existential status of the wavefunction in Everettian theory. By way of pursuing that thought, I want to discuss an idea which Wallace refers to as the *Hydra View* [8, p. 281]. And that coupled with an idea introduced by Tappenden in [26] and further developed in [27].

The Hydra view is that a physicist, and the measuring device to which s/he can indexically refer, split in idealized spin measurement contexts. The pointer evolves, via decoherence, into a superposition of showing $U$ and $D$. The two elements of the superposed pointer are both pointers which have the same mass and volume as the superposed pointer itself. Each element of the superposed pointer is to be considered as a novel type of *part*; neither a spatial part nor a temporal part but a *superpositional* part.

If the world is wavefunction it's the stuff of cats, dead or alive. But how can a cat be *indefinitely* dead or alive? The Hydra view of what happens to Schrödinger's cat when you close its idealized, causally isolated box is that it quickly evolves into a superposition of two cats, one alive and the other dead, each of which is a superpositional part of the superposed cat. Given the setup in Schrödinger's box, the quantum amplitude of the dead cat increases with time whilst that of the live cat decreases. The masses of each of the two cats remains the same, for all practical purposes, so the superposed cat always definitely has the mass of one cat, since both its superpositional parts have the same mass. Schrödinger's cat is in an indefinite dead/alive state because the two cats which are its superpositional parts are neither both dead nor both alive.

I now propose to combine this analysis of the Schrödinger cat scenario with the *concrete sets* hypothesis introduced in [26, pp. 9-10]. The aim there was to make intelligible the idea that branch weight can be identified with objective probability, which seems counterintuitive because, on the face of it, if quantum processes are deterministic then that must necessarily exclude the possibility that objective probability is involved. Tappenden has described a thought experiment designed to show that determinism and objective probability are indeed compatible. Briefly, the idea is to imagine a large but not infinite set of isomorphic 'parallel universes' in which quantum processes are stochastic and isomorphic observers in each universe are about to undertake our model spin measurement. Each observer believes that the process about to take place is stochastic and, being well informed, s/he is able to assign what s/he believes to be objective probabilities to the possible outcomes. Given the law of large numbers, what happens when the parallel measurements take place is that the set of universes partitions into two subsets where different outcomes occur, the measures of the subsets equaling the objective probabilities attributed to the possible outcomes of the parallel stochastic measurement processes.

Tappenden then argues that an alternative interpretation of the mentality of observers is possible. Instead of it being supposed that there are individual observers in each universe it is possible to interpret the situation as involving a *single* observer whose mind spans all the universes. The body of the single observer is the *set* of isomorphic doppelgangers, one in each universe. And what that single observer indexically refers to as the measuring device is the *set* of isomorphic measuring devices. That single observer is in *the very same mental state* as the original multiple observers. The mental state is individuated by its mental content, which involves the beliefs that the measurement process about to be observed is stochastic and that objective probabilities can be assigned to what are possible outcomes. The change of interpretation does not involve a change in observers' mental content, it just involves supposing that there's a single *token* of the mental content rather than multiple tokens, one in each universe.

On this alternative interpretation of the setup what happens to the single observer is that s/he *splits* into two observers each observing a different outcome because the original set of doppelgangers partitions into two subsets, each exposed to a different result because the measuring device has partitioned into two subsets. If the proposed alternative interpretation of mentality is coherent, what the though experiment demonstrates is that it's possible that an observer who thinks that a process is stochastic is *mistaken*. It's intelligible that what the observer believes to be a measuring device which will show exclusively one or another of two outcomes is *in fact* a measuring device which will split into two devices each showing a different outcome. And it's possible that what the observer takes to be objective probabilities do not attach to possibilities but rather to the *subset measures* which correspond to Everettian branch weights in the imaginary setup.

The alternative interpretation of the mind-body relation in the thought experiment involves the idea that a set of isomorphic doppelgangers instances a single observer and that objects in that observers environment are sets. The interpretation requires the concrete sets hypothesis and that entails that if there exist two ordinary, everyday cats, each with a mass of 5kg., one dead and the other alive, then the set of those two cats is also a cat. The set has a definite mass equal to the mass which the dead and live cats have in common, 5kg. And the cat which is the set is neither dead nor alive because the cats which are its elements are neither both dead nor both alive.

The concrete sets hypothesis is required for the alternative interpretation of the mind-body relation and so appears to be necessary to making intelligible the idea that what has been thought to be a stochastic process having possible outcomes with objective probabilities is in fact a branching process where co-existing branches have associated objective probabilities. So it's plausible that the concrete sets hypothesis may have a fundamental role to play in making Everettian theory intelligible. And the concrete sets hypothesis entails that *any* set of concrete objects is itself a concrete object with properties which may, or may not, be definite. Schrödinger's cat, on the Hydra view, is a superposition which has a live and a dead cat as elements (superpositional parts). What the above argument suggests is that the elements of a superposition are elements *in the set-theoretic sense*. In other words superpositional parts are not a novel type of part as suggested by the Hydra view but rather set-theoretic *elements*.

Schrödinger's cat, on the Hydra view, is the product of decoherence. When the box is closed the radioactive sample in the poison-triggering mechanism is in a superposition which decoheres into decayed and non-decayed samples, triggering the evolution of the cat into a decohered superposition of dead and live cats. But there's no reason to suppose that the physical evolution of a superposition should change its *metaphysical* constitution. In which case what physicists call the elements of *any* superposition are elements in the set-theoretic sense.

Let's now apply these ideas to the problem of wavefunction monism. Tim Maudlin has posed the problem very clearly:

> In sum, any theory whose physical ontology is a complete wavefunction monism automatically inherits a severe interpretational problem: if all there is is the wavefunction, an extremely high-dimensional object evolving in some specified way, *how does that account for the low-dimensional world of localized objects that we start off believing in, whose apparent behavior constitutes the explanandum of physics in the first place?* […] And it should be obvious that all the resources of the phrase 'configuration space' are legitimately available to a non-monist who postulates a plethora of localized particles (or strings, or whatever) in a common low-dimensional space.
>
> [27, pp. 132-33, original emphasis]

Consider the bound electron in a hydrogen atom in the light of the concrete sets hypothesis and the idea that the elements of a superposition are elements in the set-theoretic sense. At any give moment the electron is a 'point particle' with indefinite position, which is to say that it has elements at different positions in the 'electron-cloud', each element being an electron. The 'electron-cloud' surrounding the nucleus is an electron which is a set of electrons.

Imagine a field full of cats, each with a mass of 5kg. and each necessarily at a different position. The cat which is the set of those cats, according to the concrete sets hypothesis, has a definite mass of 5kg. but indefinite position. The electron which is the set of electrons in the electron-cloud is like the set of cats in the field. A natural minimal position for a point particle would be a Planck volume so for each Planck volume in the electron-cloud there is an electron which has a share of position amplitude and phase and, again, the single bound electron in the hydrogen atom, at any given moment, is the set of all those electrons. And each of those electrons will be a set of electrons with different momenta. To be sure, the picture becomes more complex if we consider the quantum wavefuncion of two or more *entangled* point particles. Now there will be two or more elements of each particle present for each Planck volume and the relation between them will embody the entanglement of the particles at that position. For an environment such as ours, where immense numbers of particles are entangled, the picture becomes extremely complex but now we have Maudlin's 'plethora of localized particles' with which to construct a metaphysics for wavefunction monism. It may seem an extravagant speculation, but perhaps I've succeeded showing that it's not unreasonable.

**7 Conclusion**

Everettian theory refuses to lie down and die, despite many attempts to kill it off. And yet there's no consensus amongst Everettian theorists as to how probability is to be understood in branching contexts nor how the metaphysics of branching itself is to be interpreted. There's also disagreement about what role wavefunction realism might, or might not, have to play in Everettian theory. I hope to have clarified some of the issues involved and to have made an intelligible suggestion as to how wavefunction monism may be compatible with Everettian theory.

In his Nobel address Max Born said:

> How does it come about then, that great scientists such as Einstein, Schrödinger and De Broglie are nevertheless dissatisfied with the situation? […] The lesson to be learned from what I have told of the origin of quantum mechanics is that probable refinements of mathematical methods will not suffice to produce a satisfactory theory, but that somewhere in our doctrine is hidden a concept, unjustified by experience, which we must eliminate to open up the road. [28]

Everett wrote:

> Arguments that the world picture presented by this theory is contradicted by experience because we are unaware of any branching process are like the criticism of the Copernican theory that the mobility of the earth as a real physical fact is incompatible with the common sense interpretation of nature because we feel no such motion. [1]

That anticipates Gertrude Anscombe's recollection of an exchange with Ludwig Wittgenstein:

> He once greeted me with the question: 'Why do people say that it was natural to think that the sun went round the earth rather than that the earth turned on its axis?'. I replied, 'I suppose, because it looked as if the sun went round the earth'. 'Well,' he asked, 'what would it have looked like if it had *looked* as if the earth turned on its axis'? [29][3]

Transposed to the Everettian context the exchange might go like this. 'Why did physicists think it natural to believe that the measurement of a superposition on the pointer basis yields a single result rather than multiple results'? Reply: 'I suppose because it looked as if a single result is yielded'. Response: 'Well, what would it have looked like if it had *looked* as if multiple results are yielded?'.

Is Born's 'hidden concept, unjustified by experience' simply the idea that physicists don't split when they observe quantum superpositions?

---

[3] Quoted by Michael Lockwood [30].